\documentclass{elsarticle}
\journal{Icarus}
\usepackage[T1]{fontenc}
\usepackage[english]{babel}
\usepackage{graphicx}
\usepackage{epstopdf} 
\usepackage{amssymb}
\usepackage{amsmath}
\usepackage{mathabx}
\usepackage{morefloats}
\usepackage{tabularx}
\usepackage{natbib}
\usepackage{url}
\newcommand{\rcom}[1]{{#1}}

\begin{document}

\begin{frontmatter}

\title{Discovering Earth's transient moons with the Large Synoptic Survey Telescope}
\author[uh]{Grigori Fedorets\corref{cor1}}
\author[uh,ltu]{Mikael Granvik}
\author[uw]{R. Lynne Jones}
\author[uw]{Mario Juri\'{c}}
\author[ifa]{Robert Jedicke}
\address[uh]{Department of Physics, Gustaf H\"{a}llstr\"{o}min katu 2, P.O. Box 64, 00014 University of 
Helsinki, Finland}
\address[ltu]{Division of Space Technology, Lule\aa{} University of Technology, Box 848, S-98128 Kiruna, Sweden}
\address[uw]{DIRAC Institute, Department of Astronomy, University of Washington, 
Box 351580 Seattle, WA 98195-1580, U.S.A.}
\address[ifa]{Institute for Astronomy, University of Hawaii, 2680 
Woodlawn Dr, Honolulu, HI 96822, U.S.A.}

\cortext[cor1]{Corresponding author, fedorets@iki.fi}

\begin{abstract}


Earth's temporarily-captured orbiters (TCOs) are a sub-population of near-Earth objects (NEOs). TCOs can provide constraints for NEO population models in the 1--10-metre-diameter range, and they are outstanding targets for in situ exploration of asteroids due to a low requirement on $\Delta v$. So far there has only been a single serendipitous discovery of a TCO. Here we assess in detail the possibility of their discovery with the upcoming Large Synoptic Survey Telescope (LSST), previously identified as the primary facility for such discoveries. We simulated observations of TCOs by combining a synthetic TCO population with an LSST survey simulation. We then assessed the detection rates, detection linking and orbit computation, and sources for confusion. Typical velocities of detectable TCOs will range from 1$^{\circ}$/day to 50$^{\circ}$/day, and typical apparent $V$ magnitudes from 21 to 23. Potentially-hazardous asteroids have observational characteristics similar to TCOs, but the two populations can be distinguished based on their orbits with LSST data alone. We predict that a TCO can be discovered once every year with the baseline moving-object processing system (MOPS). The rate can be increased to one TCO discovery every two months if tools complementary to the baseline MOPS are developed for the specific purpose of discovering these objects.  
\end{abstract}

\end{frontmatter}

\section{Introduction}

Following the discovery and characterisation of Earth's temporary-captured satellite 2006 RH$_{120}$ \citep{kwiatkowski2009},  \citet{granvik2012} were the first to propose that there is a population of small asteroids in orbit around the Earth at any given time. Further evidence of temporarily-captured satellites come from, e.g., backwards orbit integration of the asteroid 1991 VG \citep{delafuentemarcos2018}, the discovery of the meteor EN130114 originating from a geocentric orbit \citep{clark2016}, and a candidate temporarily-captured satellite detected by the Space Surveillance Telescope \citep{lue2019}.  Natural temporarily-captured satellites include both temporarily-captured orbiters (TCOs), that make at least one equivalent of a revolution around the Earth while being captured, and temporarily-captured flybys (TCFs), that make less than one equivalent of a revolution while being captured. We will omit TCFs from the following analysis due to their short capture duration which makes them less interesting targets for follow-up observations and space missions compared to TCOs. A recent analysis by \citet{fedorets2017} yielded 75 cm as the diameter of the largest TCO at any instant in time. 

The definition for a temporary capture is somewhat ambiguous \citep{fedorets2017,urrutxua2017,jedicke2018}. In this work we use the definition of
\citet{fedorets2017}, which requires that the geocentric capture needs to fulfil the following criteria:
\begin{itemize}
	\item $e_{\Earth} < 1$.
	\item The asteroid is within 3 Hill radii from the Earth.
	\item The asteroid approaches the Earth to within the Hill radius distance during the capture.
\end{itemize}

The interest towards Earth's temporarily-captured sa\-tel\-li\-tes is two-fold. First, they are potential targets for rendezvous missions with CubeSat-sized spacecraft or, more speculatively, even asteroid-return missions with larger spa\-ce\-craft. Second, the systematic discovery of bodies with diameters ranging from decimeters to a decameter would constrain the size-frequency distribution (SFD) of near-Earth objects (NEOs) in that size range and resolve the existing discrepancies between the different techniques used in different size ranges. The larger end of the NEO SFD is based on telescopic observations of NEOs \citep{rabinowitz2000,harris2015,granvik2016a,tricarico2017} whereas the smaller end is based on an analysis of bolide data \citep{brown2002,brown2013}. TCOs are the closest steady-state small-body population to the Earth, and can provide the most comprehensive distribution of small, yet observable asteroids. Therefore, constructing a population model of TCOs could be used to solve the existing contradictions between different models in the 0.1--10-m-diameter range for the entire NEO population.

\begin{figure}%
\includegraphics[width=\columnwidth]{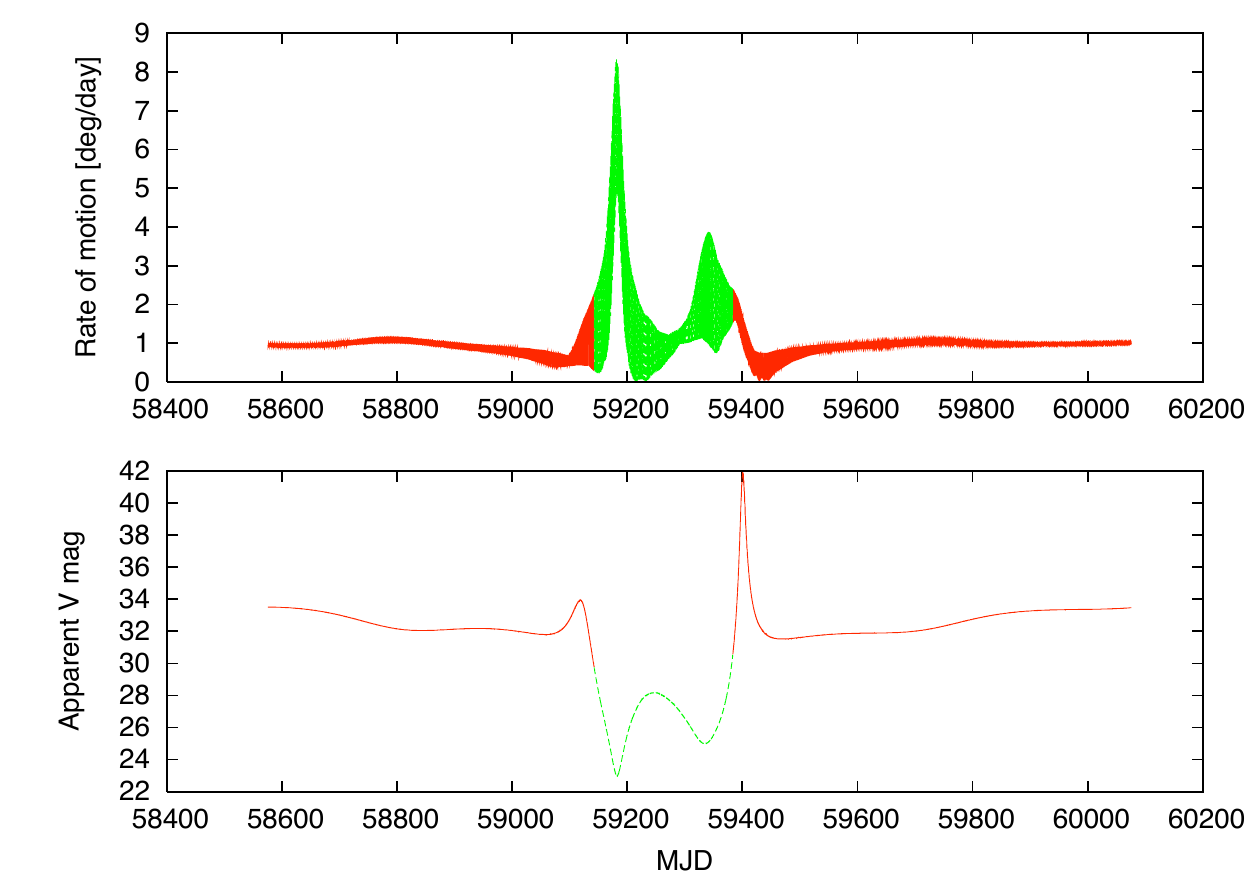}%
\caption{A typical example of the combination of topocentric rate of motion and apparent brightness illustrating the difficulty of TCO observations. \emph{Top image:} rate of motion as a function of time. \emph{Bottom image:} apparent magnitude as a function of time. The green line represents the period during which the object is captured as a TCO whereas the red line represents the time evolution of the parameters when the object is not captured.}%
\label{fig:ratemotion}%
\end{figure}

The problem with discovering TCOs is that the largest of them, and also the ones that can be observed, are among the smallest known asteroids and that, in turn, typically makes them very faint. \rcom{The diameter of 2006 RH$_{120}$ was 3 metres. In that size range, objects are expected to become captured on the order of once every ten years.} In addition, TCOs tend to move relatively fast when available for detection in terms of their apparent brightness (see Fig.~\ref{fig:ratemotion}). 

\citet{bolin2014} investigated various options for discovering TCOs with current and upcoming facilities. Out of the  instruments available in 2019, the HyperSuprimeCam at the Subaru Telescope was deemed the best solution.  If a TCO were to be discovered with HyperSuprimeCam or any other instrument, it would be larger than the largest object in a steady-state population. Therefore, such a discovery would be of serendipitous nature. The Space Surveillance Telescope (SST) has detected the smallest known NEO \citep{lue2019}, but its operational mode is optimised for the discovery of very close artificial bodies whereas \rcom{ TCOs  usually get} confused with more distant artificial objects such as those launched beyond the geostationary orbits (e.g., interplanetary or lunar launches, or missions to the $L_2$ region). TCOs are recognised as worthy follow-up candidates in the asteroid community, and potential candidates are regularly flagged by automatic short-term alert systems such as Scout \citep{farnocchia2015} and NEORANGER \citep{solin2018}. For example, in 2017 Scout identified two TCO candidates, but subsequent astrometric observations showed that both of them were of artificial origin based on their high area-to-mass ratios. \rcom{For one of these objects, 2018 AV$_2$, a month's worth of astrometric observations were required to rule out the natural origin \citep{jedicke2018}.}

The upcoming Large Synoptic Survey Telescope \citep[LSST,][]{ivezic2019} was identified as the most promising facility for sustained TCO discovery among existing and upcoming facilities \citep{bolin2014}. The following combination of factors make LSST particularly suitable for discovering TCOs:  the light-gathering area of the mirror is suitable for detecting faint objects, the relatively short exposure time compared to pixel size reduce trailing losses, the wide field of view allow for a significant part of the visible sky to be observed each night, and the observational cadence allows for the identification of moving objects. 

Despite these capabilities, TCOs are challenging objects even for an advanced system such as LSST. Discovering TCOs has nevertheless been identified as one of the scientific goals for NEO studies in the LSST solar system science roadmap \citep{schwamb2018}.

The invention of space flight has put thousands of artificial objects on geocentric orbits that may be a source of confusion when attempting to identify TCOs. The vast majority of artificial objects are on low-Earth orbits (LEO) and on geosynchronous equatorial orbits (GEO), which differ from typical orbit of TCOs. However, there is a number of artificial objects with distant orbits that can be confused with natural objects. The Minor Planet Center (MPC) lists them on the Distant Artificial Satellites Observation Page\footnote{https://www.minorplanetcenter.net/iau/artsats/artsats.html}. If these objects cannot be attributed to any launch, the best way to distinguish between a natural or artificial origin for these bodies is their area-to-mass ratio. Constraining the area-to-mass ratio often requires extensive astrometric coverage, and this leaves the status of the object in ambiguity for extended periods of time. 

Proving that Earth's temporarily-captured natural satellites can be discovered using LSST data will increase the scientific interest towards them --- in particular, as a population of asteroids in near-Earth space that are easily accessible by future space missions. In this work we will answer the following questions:
\begin{itemize}
	\item How frequently can TCOs be discovered?
	\item What are their typical magnitudes and velocities when they are detected?
	\item Can observations of TCOs be identified in the LSST data swarm?
	\item Can observations of TCOs be linked from one night to the next night?
	\item Is it possible to distinguish them from other close-approaching asteroids and artificial objects?
\end{itemize}

In this work, we assess the probability of routine detection of TCOs with diameters ranging from 10 cm to 4 m using synthetic TCOs, our current best guess for the cadence of LSST observations and the performance of the LSST data processing pipeline. We also assess the steps required to ensure that follow-up observations can be triggered rapidly. We describe LSST's capabilities in Sect.~\ref{s:lsst-c}, introduce the constraints and design of the survey simulation in Sect.~\ref{s:susi}, present the results, linking simulation, and distinguishing TCOs from the possible sources for confusion, as well as related discussion in Sect.~\ref{s:randd}, and offer our conclusions in Sect.~\ref{s:c}.

\section{Large Synoptic Survey Telescope} \label{s:lsst-c}

Encouraged by the potential of large surveys, and at the same time acknowledging the constraints of the previous surveys, a call for a new ground-based synoptic survey emerged in the early 2000s \citep[for details, we refer the reader to][and references therein]{ivezic2019}. This eventually led to the construction of LSST. The four science themes of LSST are:
\begin{enumerate}
	\item Probing Dark Energy and Dark Matter.
	\item Taking an Inventory of the Solar System.
	\item Exploring the Transient Optical Sky.
	\item Mapping the Milky Way.
\end{enumerate}
The design of LSST is therefore striving to accommodate all four science themes and specific science goals within each theme. Here we remind the reader of the aspects relevant for solar system studies and, in particular, for TCOs. The major drive for the solar system objectives is the asteroid impact threat, which is concretised in a mandate by the Congress of the United States of America  to discover 90\% of potentially-hazardous asteroids (PHAs) with a diameter greater than 140 metres. A detailed presentation of the science goals pertaining to solar system studies in general are described by \citet{jones2009} and \citet{schwamb2018}, to discovering asteroids with LSST by \citet{jones2015} and \citet{jones2017}, and particularly to discovering NEOs by \citet{jones2018}. NEO discovery with LSST has also been independently studied by \citet{grav2016} and \citet{veres2017a,veres2017b}.

LSST will become operational in the early 2020's and is expected to cover the available sky from its location in Cerro Pach\'{o}n in northern Chile every few nights for the duration of 10 years. LSST combines in an unprecedented way its large aperture of 8.4 metres (6.5 metres effective) with the wide-field optics allowing for a 9.6-square-degree field of view recorded with the 3.2 Gigapixel camera with 10 $\mu$m pixels, automatic processing of extended objects, and an expected limiting magnitude of $r=24.6$ (dark sky, zenith) for non-moving point sources \citep{jones2009}. Each night, 10\,000 square degrees of the sky will be covered, yielding 15--20 terabytes worth of data. For comparison, an LSST data set collected during a single night is comparable to a decade of Sloan Digital Sky Survey (SDSS) data releases.

To accommodate the requirement to detect moving objects, the LSST cadence will include two separate visits to the same field during a night, followed by a revisit within a few days. The debate for the final observational cadence is still ongoing. In this work we use one of the options for the cadence, called \textit{kraken\_2026} \citep{boberg2018}, which in early 2019 was the top-running candidate for the operational baseline cadence. An alternative to the \textit{kraken\_2026} cadence is, for example, the so-called rolling cadence which concentrates on a single region of the sky at a time rather than spreading the observations across the entire visible sky every few days. Rolling cadence is advantageous for follow-up of closely approaching, and hence fast-moving NEO flybys and TCOs.

\rcom{While depending somewhat on the seeing (mean seeing value on-site is expected to be 0.8$^{\prime \prime}$, which is the trail length of an asteroid moving with a velocity of 0.64$^{\circ}$/day), all}  objects that move faster than 1$^{\circ}$/day will be trailed in LSST images \citep{veres2017a}. Two trailed detections in a 15-90 minute window will identify the object as an NEO candidate. The trailed sources will be measured using point-spread-function (PSF) photometry and fit using a trailed point source model \citep{juric2018,schwamb2018}. It should be noted that the resulting detections will also include false positives.

In the current baseline approach the data is processed through LSST's Moving Object Processing System \citep[MOPS; see ][for a description of the original Pan-STARRS MOPS]{denneau2013} using the \emph{findtracklets} and the \emph{linktracklets} algorithms by \citet{kubica2007} followed by initial orbit computation. Detections in difference images will first be linked into single-night \rcom{batches of observations colloquially known as \emph{tracklets}} using the \emph{findtracklets} algorithm. The \emph{linktracklets} algorithm will then attempt to link groups of three tracklets obtained within the span of 15 nights into \emph{tracks} by using a polynomial, typically of first or second order. Initial orbit computation will then be used to remove tracks  that cannot be described with a physically-meaningful trajectory \citep{veres2017b}.

\section{Survey simulation} \label{s:susi}

\subsection{Orbits} \label{s:simorb}

As input for the LSST survey simulation we used the 20,265 synthetic TCOs generated and described by \citet{fedorets2017}. \rcom{In summary, they first sampled the phase space of heliocentric Keplerian elements by randomly drawing orbits from a volume containing orbits similar to that of the Earth's orbit: 0.87 au < $a_{\Sun}$ < 1.15au; $e_{\Sun}$ < 0.12; $i_{\Sun}$ < 2.5$^{\circ}$ whereas the longitude of ascending node $\Omega_{\Sun}$, argument of perihelion $\omega_{\Sun}$, and mean anomaly $M_{\Sun}$ were randomly drawn from a uniform distribution of angles spanning $2\pi$. The initial geocentric distance was required to be between 4 and 5 Hill radii (i.e., 0.04--0.05 au). The sampling resulted in almost 10 million test asteroids that were then propagated through the Earth-Moon system and out of these 20,265 turned out to be TCOs.}

\rcom{To estimate the size-frequency distribution of the steady-state population, \citet{fedorets2017} assumed that the flux of asteroids from the NEO population to the intermediate source population, that is, the phase-space volume containing capturable orbits (see definition above), is proportional to the flux from the intermediate source population to the TCO population thereby following the technique described in \citet{granvik2012}. The lower boundary of the steady-state population was anchored to the predicted flux into the intermediate source population at $H=25$, based on the NEO model by \citet{granvik2016a}. The $H$ distribution for TCOs of smaller sizes is an extrapolation from that base value with the slope of the distribution obtained from bolide data \citep{brown2002}.}

The sample of synthetic TCOs is spread over the Me\-to\-nic cycle --- a 19-year cycle after which the Sun-Earth-Moon geometry is repeated. \rcom{\citet{fedorets2017} chose MJD 54466.0 (January 1st 2008) as the starting epoch for the Metonic cycle and their orbital integrations. Since we are not redoing the integrations here, this is a limitation that we have to live with in the analysis that follows.  An absolute magnitude $H$ was randomly assigned to each synthetic TCO following the $H$ distribution derived in \citet{fedorets2017}.} The size of the largest body in the sample was selected to match the largest TCO expected to be captured once every 19 years, i.e., $H \approx 29.8$ or a diameter of about 4 m assuming a geometric albedo $p_V=0.14$. \rcom{The smallest object in the population is on the order of 10 cm.}

\subsection{Pointing}

Ephemerides of the synthetic TCOs were then matched against the \textit{kraken\_2026} pointing simulation of LSST \citep{boberg2018} using the LSST Operations Simulation Tool \citep{delgado2014} and analysed using the LSST Metrics Analytics Framework tool \citep{jones2014}.
  
The pointing simulation takes into account the various LSST survey modes. The main survey mode, the so-called wide-fast-deep survey, covers the southern sky between declinations 0$^{\circ}$ and -60$^{\circ}$ and consumes around 90\% of the total observing time. In addition, the north ecliptic spur is included, adding regions of the northern sky to ensure that the entire ecliptic plane is covered to $\pm10$ degrees.

\begin{figure}%
\includegraphics[width=\columnwidth]{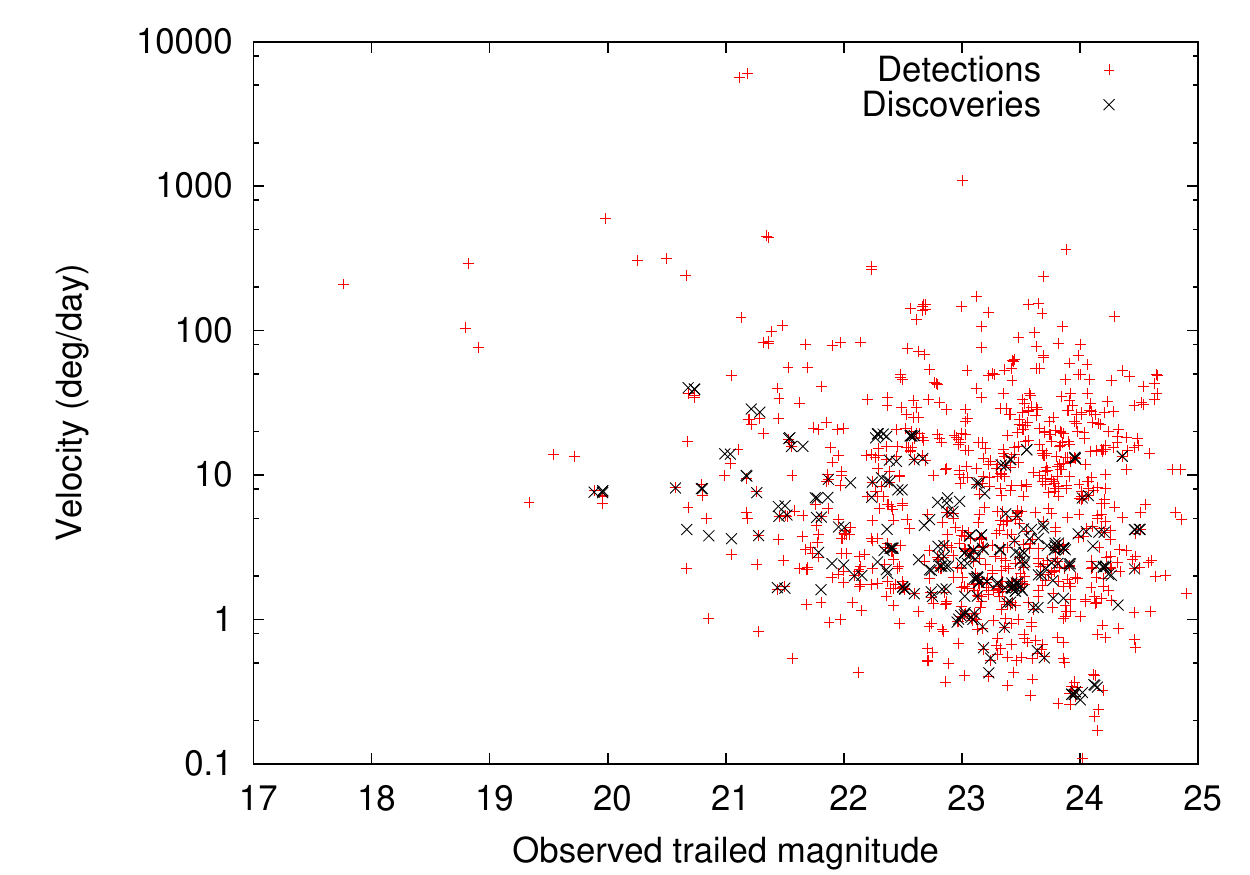}%
\caption{The distribution of the observed trailed magnitudes of all TCO detections and points contributing to discoveries as a function of sky velocity.}
\label{fig:vvvel}%
\end{figure}

\begin{figure}%
\includegraphics[width=\columnwidth]{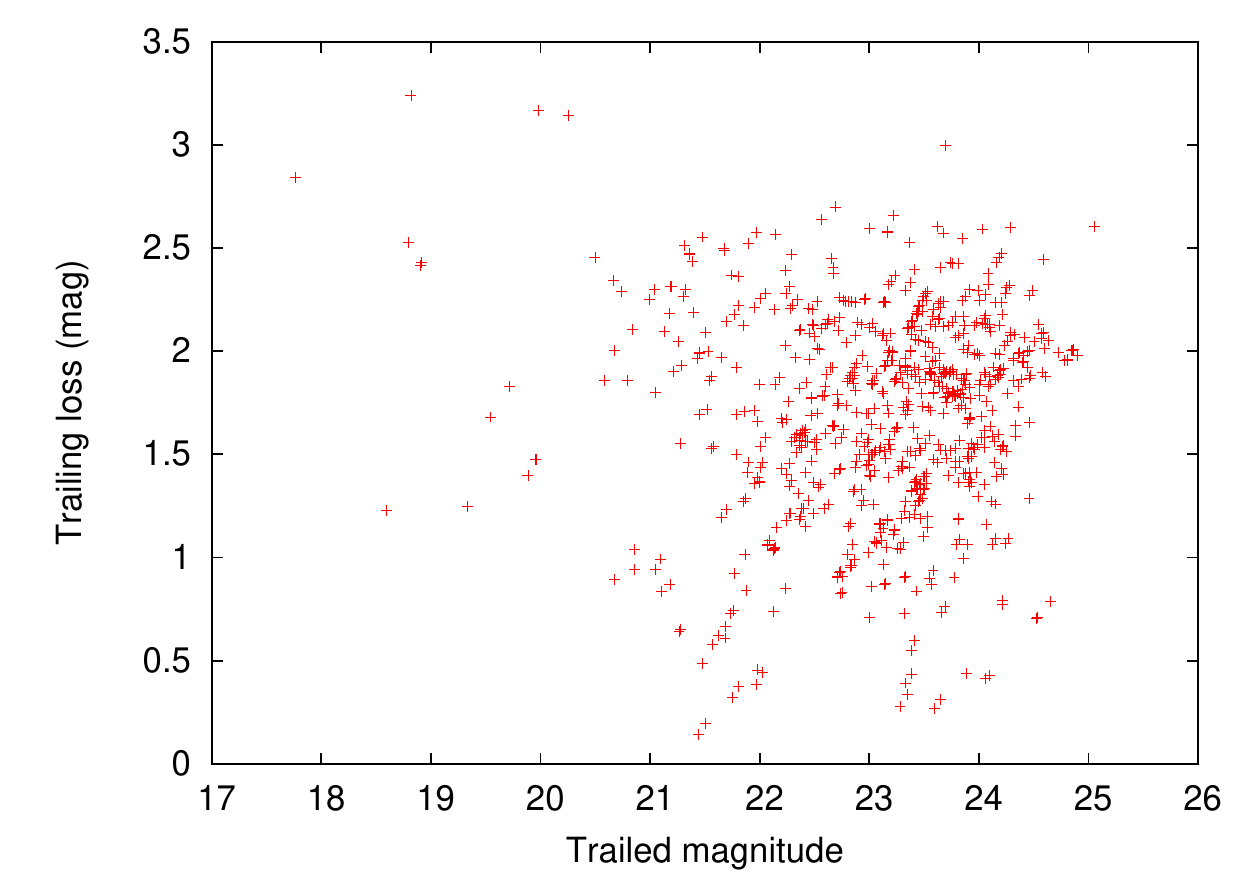}%
\caption{The  observed trailed magnitude as a function of the trailing loss of magnitude for all TCO detections.}%
\label{fig:trlo}%
\end{figure}

\section{Results and discussion} \label{s:randd}

\subsection{Baseline survey performance} \label{ss:psor}

\begin{figure}%
\includegraphics[width=\columnwidth]{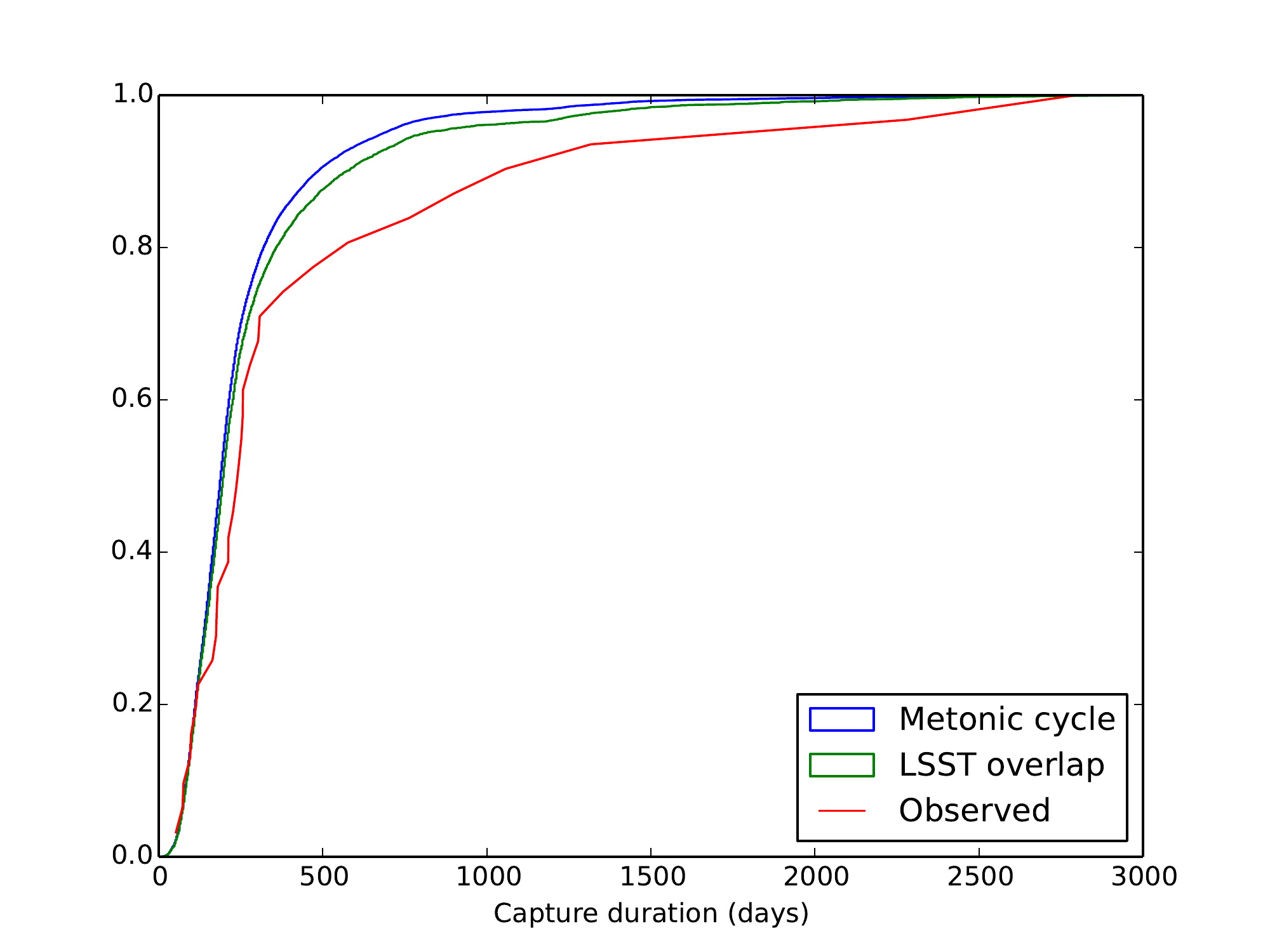}%
\caption{Comparison of normalised cumulative distribution of capture durations of TCOs over an entire Metonic cycle, for the simulation overlap with LSST and for all detected objects.}%
\label{fig:cumhist}%
\end{figure}

The TCO simulation and the LSST pointing simulation do not coincide perfectly, but do overlap over a period of 4.25 years (2022--2027). Our results are based on synthetic TCOs from that period. \rcom{One can question whether the results for a 4.25-year survey can be used to assess the discovery rate during a 10-year survey, because one might think that the discovery likelihood increases with a longer capture duration. To this end we compare the normalised cumulative capture-duration distributions for the entire TCO sample, for TCOs overlapping with the LSST pointing simulation, and for TCOs discovered in the LSST survey simulation (see Fig.~\ref{fig:cumhist}). A two-sample Kolmogorov-Smirnov statistical test for the distribution pairs (entire versus overlapping, and overlapping versus discovered) shows that we cannot reject the null hypothesis that the distributions are drawn from the same population at the level of 5\%. Hence there is no indication that our limited survey simulation would be biased against finding TCOs with either long or short capture durations.} 

The baseline assumption for the LSST is that the detections will be linked into tracklets and tracks with existing MOPS software. The basic criterion for a MOPS discovery is that a pair of visits, separated by more than 15 and less than 90 minutes, occurs during any three nights during a 15-night observing window \citep{jones2018} \rcom{requiring six separate observations}. Only three objects fulfilled this criterion in 4.25 years. While nearly all objects appear in the field of view, they are typically too faint to overcome the detection threshold. The results of the survey simulation are summarised in Table~\ref{tab:bas}.

\begin{table}%
\caption{Summary of results from the baseline survey simulation, and initial orbit determination for objects with three or more observations when using an alternative approach.}
\label{tab:bas}
\begin{center}
\begin{tabular}{lc}
\hline
Total number of TCOs                                    & 20\,265 \\
Number of TCOs during simulation overlap                  & 4\,554 \\
Number of TCOs that appear in the survey area             & 4\,551 \\
Number of TCOs with at least & \\
\quad - one & 161 \\
\quad - two & 132 \\
\quad - three & 73 \\
 detections & \\
Number of objects with cadence  &    3\\
allowing discovery with MOPS       &  \\
\hline
Objects with $S_\mathrm{geo}=100\%$ &      32  \\
Objects with $40\%<S_\mathrm{geo}<99\%$ &    10 \\
Objects with $25\%<S_\mathrm{geo}<40\%$ & 2 \\
\hline
\end{tabular}
\end{center}
\end{table}

\citet{bolin2014} predicted that LSST would discover around 1.5 TCOs/month, but with difficulties for follow-up. Our baseline estimate is more conservative, being closer to 1 TCO/year. One source for the discrepancy is due to the updated steady-state estimate from \citet{fedorets2017} compared to \citet{granvik2012}, where the number of TCOs is increased by 10\%. Here we also consider a detailed survey simulation whereas \citet{bolin2014} derived their estimates using average TCO properties as well as simplified cadence and performance characteristics for the LSST. Hence it has always been clear that the prediction by \citet{bolin2014} provides an upper limit for the number of expected TCO discoveries and more accurate simulations will decrease that number. \rcom{Our results are in line with predictions by \citet{bolin2014} when considering the number of objects with three or more observations.}

The distribution of the observed trailed magnitudes as a function of velocity shows a distinct absence of slow bright objects (Fig.~\ref{fig:vvvel}). The brightest detections occur during very close approaches, when also the velocities are highest. \rcom{The effect of trailing loss is taken into account in the calculation of the apparent brightness (Fig.~\ref{fig:trlo}). The typical brightness loss in trailing is between 1 and 2.5 magnitudes.} The distribution of observed trailed magnitudes near the threshold limit of $r=24.7$ is uniform as a function of trailing loss.

TCOs have a tendency of being detected  at the direction of the opposition (Fig.~\ref{fig:where}).  Therefore, increasing the number of observations towards the direction of the opposition would be beneficial for increasing the chances of detecting TCOs. This could be achieved with the rolling cadence in the direction of the opposition, which is also advantageous for the detection of other Solar system objects.

\subsection{Increasing the TCO discovery rate} \label{ss:idr}

MOPS requires observations on three nights during a 15-night period. However, TCOs may appear only for a pair of consecutive or close nights before becoming too faint to be detected. Therefore, we analysed the performance of an alternative approach to moving-object detection for objects with at least three observations. The proposed approach may be used as a starting point for a future pipeline dedicated to processing detections of fast-moving targets such as TCOs. We define the geocentric score $S_\mathrm{geo}$, as the weighted fraction of sample orbits that are elliptic in a geocentric frame at the mid-epoch of the observations. 

As a first step, we selected all TCOs with at least two detections. \rcom{We} performed initial orbit computation with statistical ranging \citep{virtanen2001} implemented in the OpenOrb software package \citep{granvik2009}. We continued the analysis by identifying objects that had at least one additional detection available within 15 days from the observation date of the tracklet. We then added one detection at a time and redid the orbit computation until we reached $S_\mathrm{geo}=100\%$ or there were no more detections available.

If ranging failed to produce an orbital solution, we solved the orbits using the Levenberg-Marquardt non-linear least-squares method \citep{levenberg1944,marquardt1963}. The we randomly sampled the hyperellipsoid defined by the covariance matrix of the least-squares solution. The final posterior distribution in the six-dimensional orbital-element phase space was then obtained with Markov-chain Monte Carlo (MCMC) and a proposal based on differences between the covariance-sampled orbits \citep{muinonen2012}. We assessed $S_\mathrm{geo}$ using the final posterior distribution. 

Using the orbital solution we then computed the geocentric score.  None of the single-night observation cases had 100\% scores: a typical case varied between 0.1\% and 2\%. The two best single-night cases yielded $S_\mathrm{geo}$ of 33\% and 38\%. These are the objects which have three detections on one night as well as very high sky velocities. Their $S_\mathrm{geo}$ is one or two orders of magnitude higher than for a typical single-night pair of detections. They are the extreme cases of TCOs which may be considered for follow-up observations.

The analysis resulted in 32 synthetic TCOs (or about 44\% of all the TCOs that are detected at least three times) obtaining $S_\mathrm{geo}=100\%$. 25 of these results were calculated with ranging, and 7 with MCMC in orbital elements. Out of the remaining 41 synthetic TCOs 12 (16\%) have $25\%<S_\mathrm{geo}<99\%$. Two of these objects are the aforementioned single-night cases, and  five have $S_\mathrm{geo}>98\%$. The sampled Keplerian orbit distribution of these objects is typically best described as being in the so-called orbital phase transition \citep{virtanen2005}, where the resulting Keplerian orbit distribution is nearly, but not entirely Gaussian. An orbital solution was produced for all but one object with at least three detections within 15 nights.

\begin{figure}%
	\includegraphics[width=0.7\columnwidth,angle=-90]{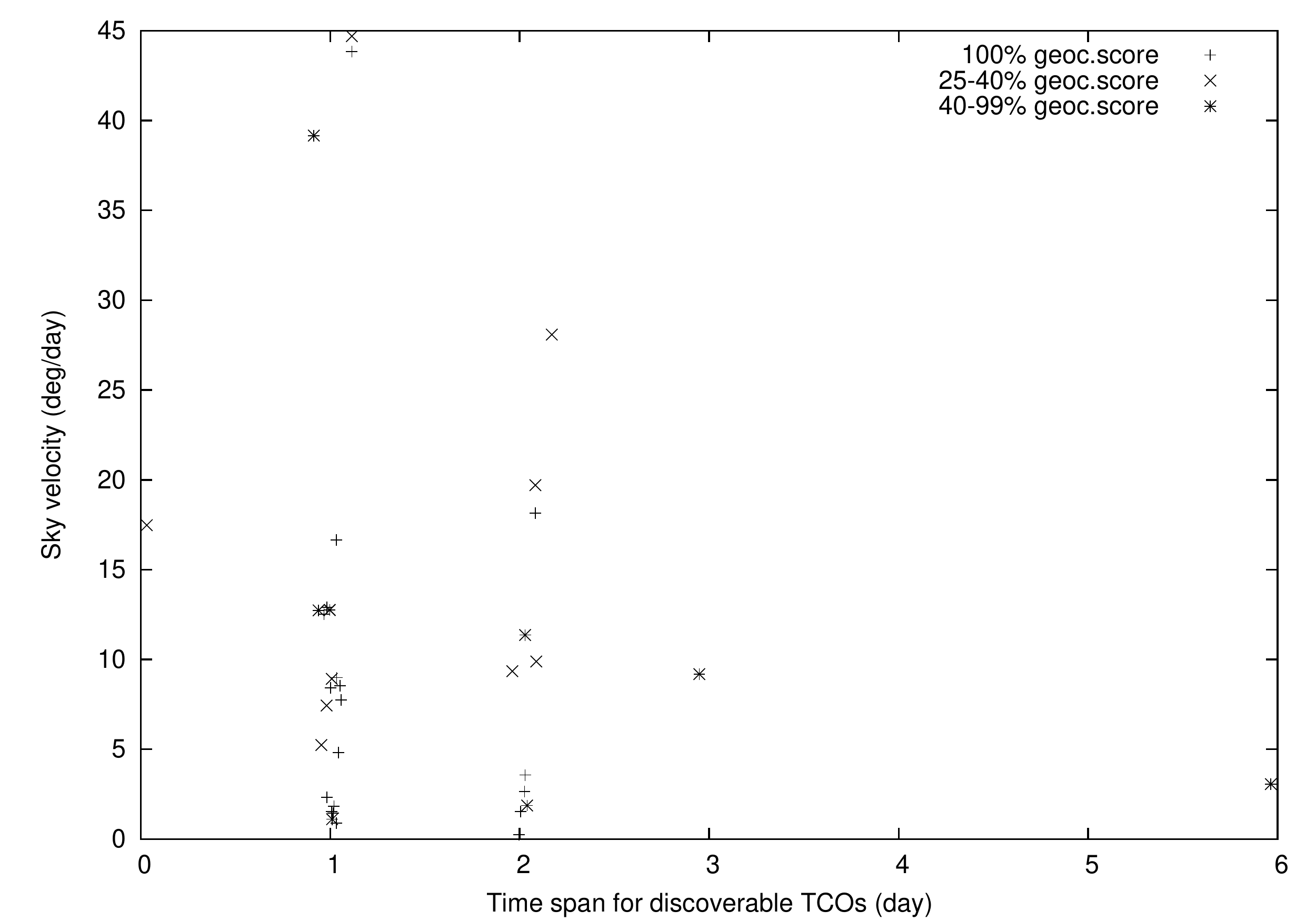}%
	\caption{Typical velocities and time spans for discoverable TCOs. A single outlier with time interval 0.03 days and velocity 90$^\circ$/day is omitted from the plot as an outlier.}%
\label{fig:tvvel}%
\end{figure}

\begin{figure*}%
\includegraphics[width=\textwidth]{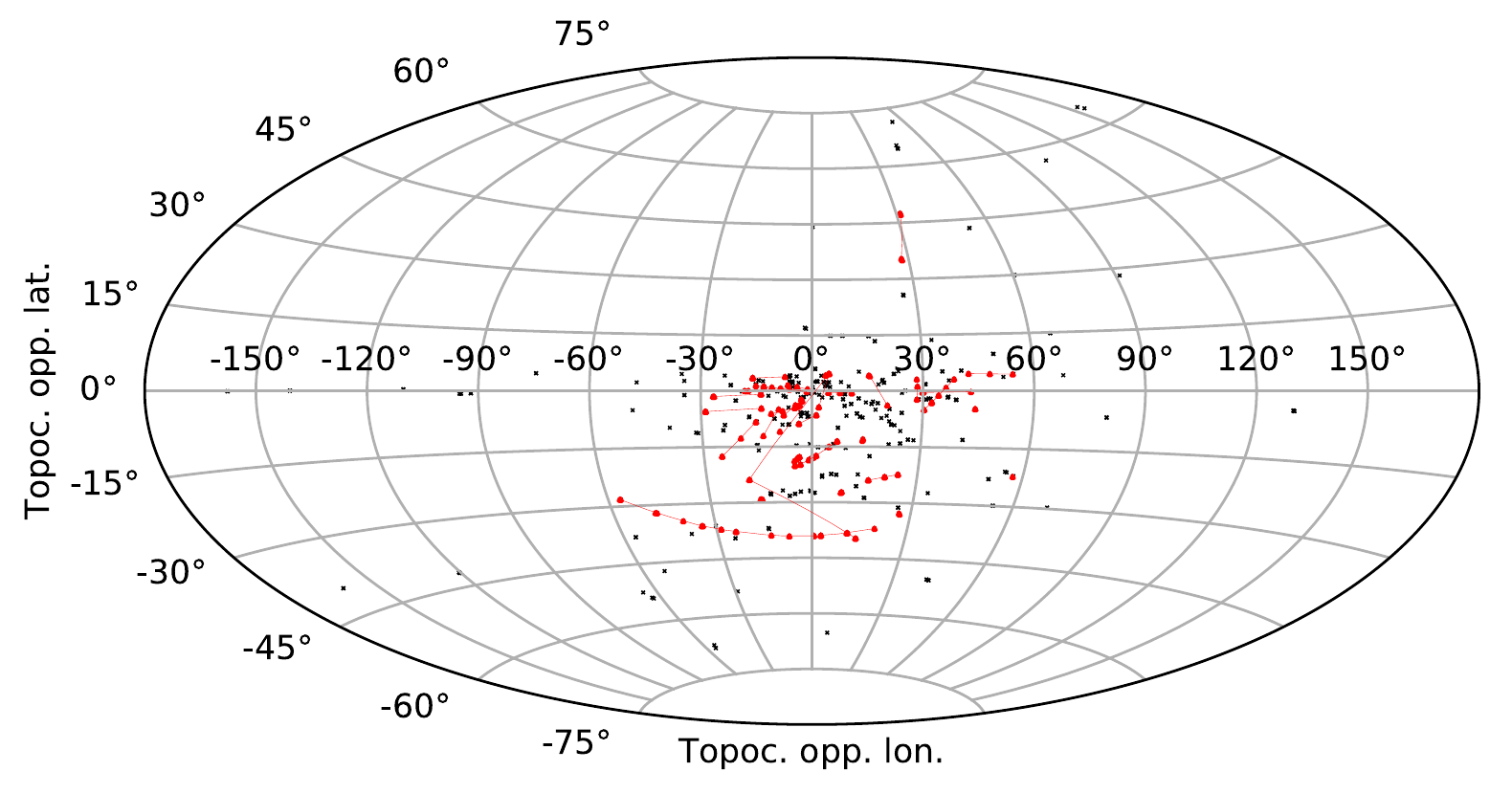}%
\caption{The distribution of TCO detections and discoveries as a function of their respective topocentric opposition-centric longitude (abscissa) and latitude (ordinate) in the Aitoff projection. Black crosses correspond to individual detections, and red triangles to detections contributing to discoveries. The detections contributing to discoveries are connected by the respective thin red lines.}%
\label{fig:where}%
\end{figure*}

\rcom{The instantaneous velocities of detected TCOs reach above 1000$^{\circ}$/day, but discoverable TCOs (i.e., objects for which $S_\mathrm{geo}>40\%)$ are concentrated towards the slower end, and do not exceed 50$^{\circ}$/day (Figs.~\ref{fig:vvvel} and \ref{fig:tvvel}). The typical interval between detections leading to discovery are one or two nights, but it can be as long as six nights (Fig.~\ref{fig:tvvel}). Mostly same fields are observed on different nights, hence the intervals are close to integer days. The distribution of discoverable TCOs is uniform throughout the detected positions, which are, in turn, concentrated towards the direction of the opposition  (Fig.~\ref{fig:where}). }

\begin{figure}%
\includegraphics[width=\columnwidth]{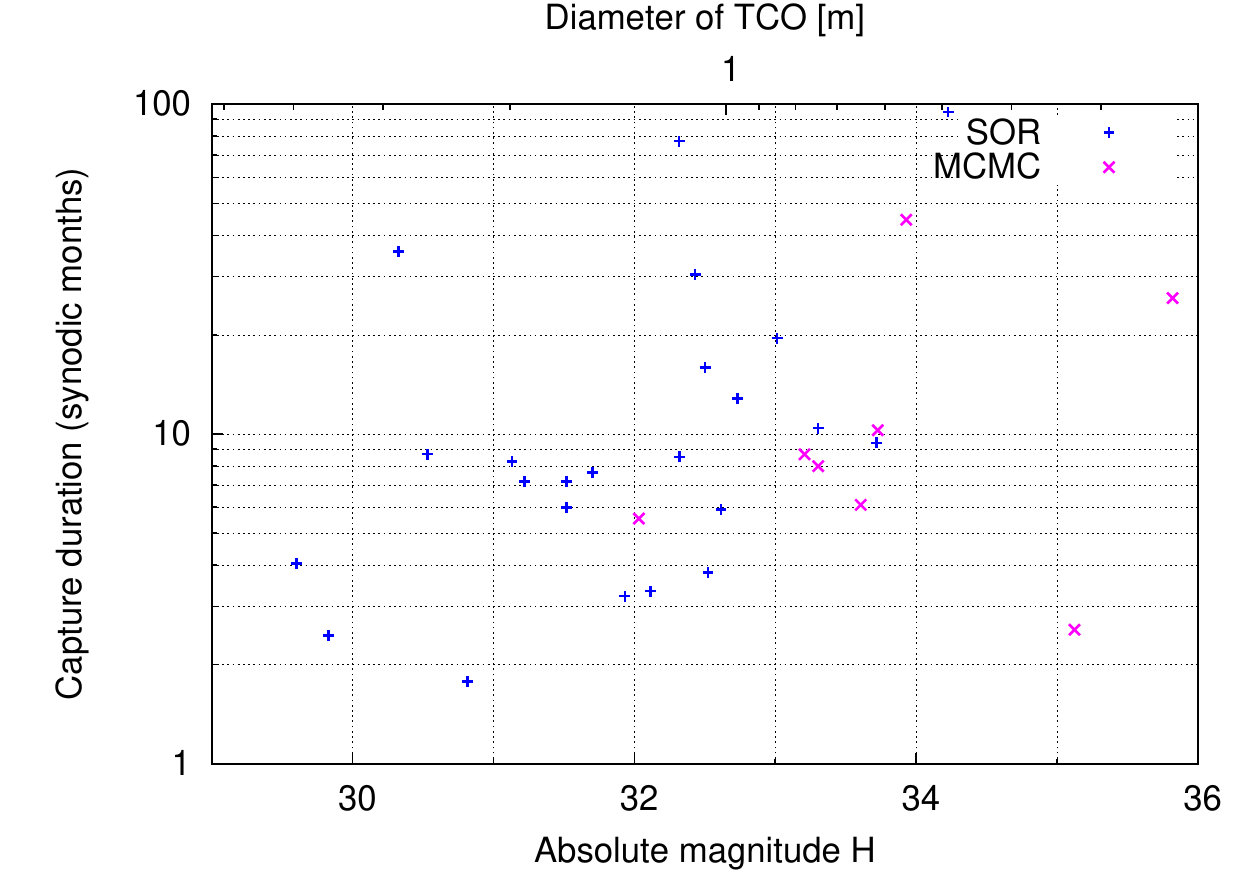}%
\caption{The relation between the absolute magnitude $H$ of TCOs with $S_\mathrm{geo}=100\%$, and their capture duration. The plus signs correspond to orbits obtained with statistical orbital ranging, and crosses to orbits obtained with the MCMC in orbital element space.}%
\label{fig:hcapdur}%
\end{figure}

The numbers from the simulations, spread over the 4.25-year simulation would suggest that on average seven TCOs with $S_\mathrm{geo}=100\%$ can be detected annually with LSST when following the \textit{kraken\_2026} cadence. Hence the discovery rate can potentially increase by an order of magnitude if an alternative to the baseline MOPS is considered for the processing. The discovered TCOs are among the largest ones in the simulated population irrespective of their capture duration (Fig.~\ref{fig:hcapdur}). In other words, size is more important than capture duration for TCO discovery.

\begin{figure}%
\includegraphics[width=\columnwidth]{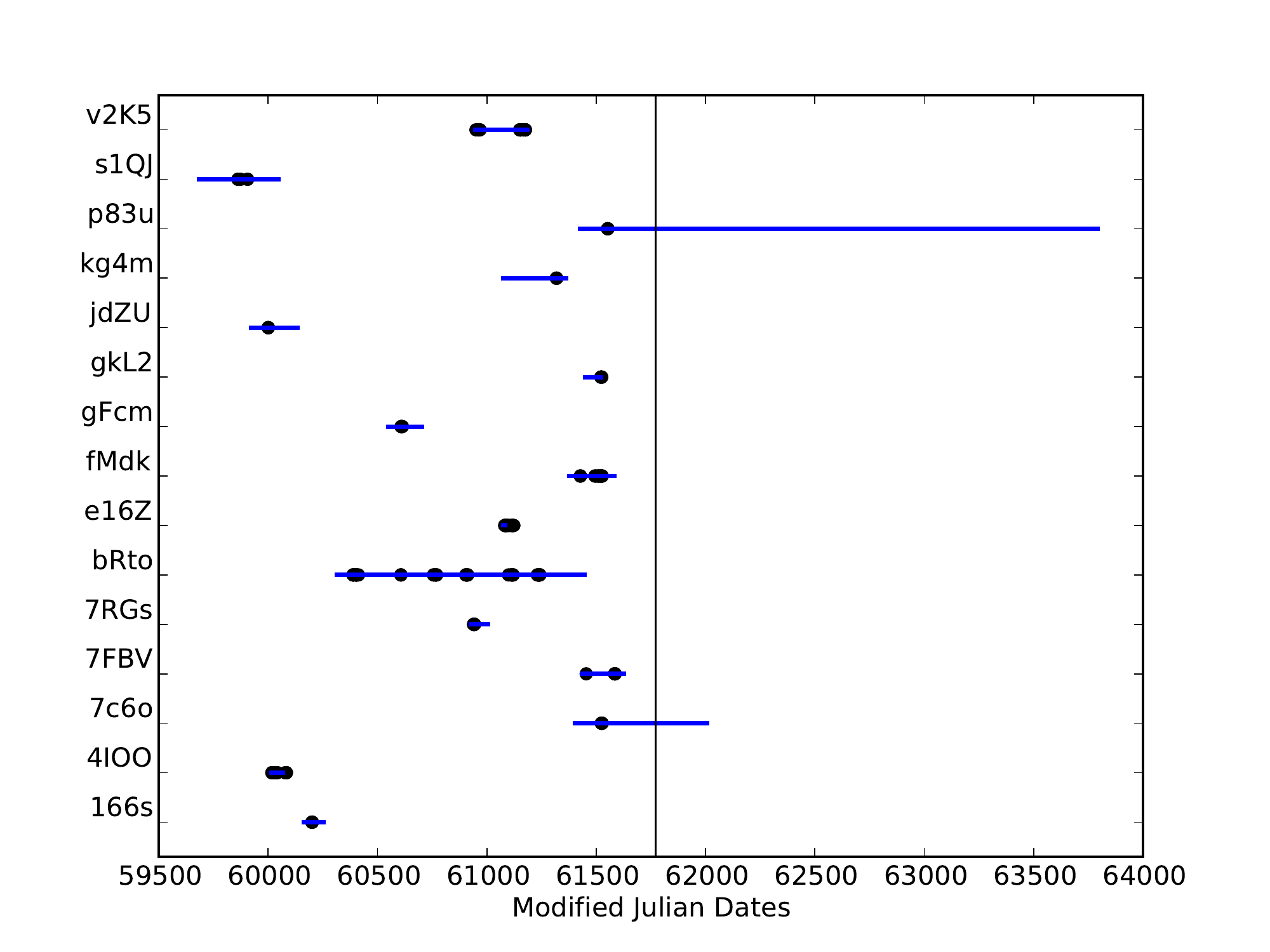}%
\caption{The detectability of a sample of discoverable TCOs during their capture. The blue thin lines represent the duration of the geocentric capture, and the black dots designate the epochs when the TCO was detected. The black vertical line represents the end of the TCO population simulation, after which no new objects appear in the survey simulation.}%
\label{fig:lineplot}%
\end{figure}

\begin{figure}%
\includegraphics[width=\columnwidth]{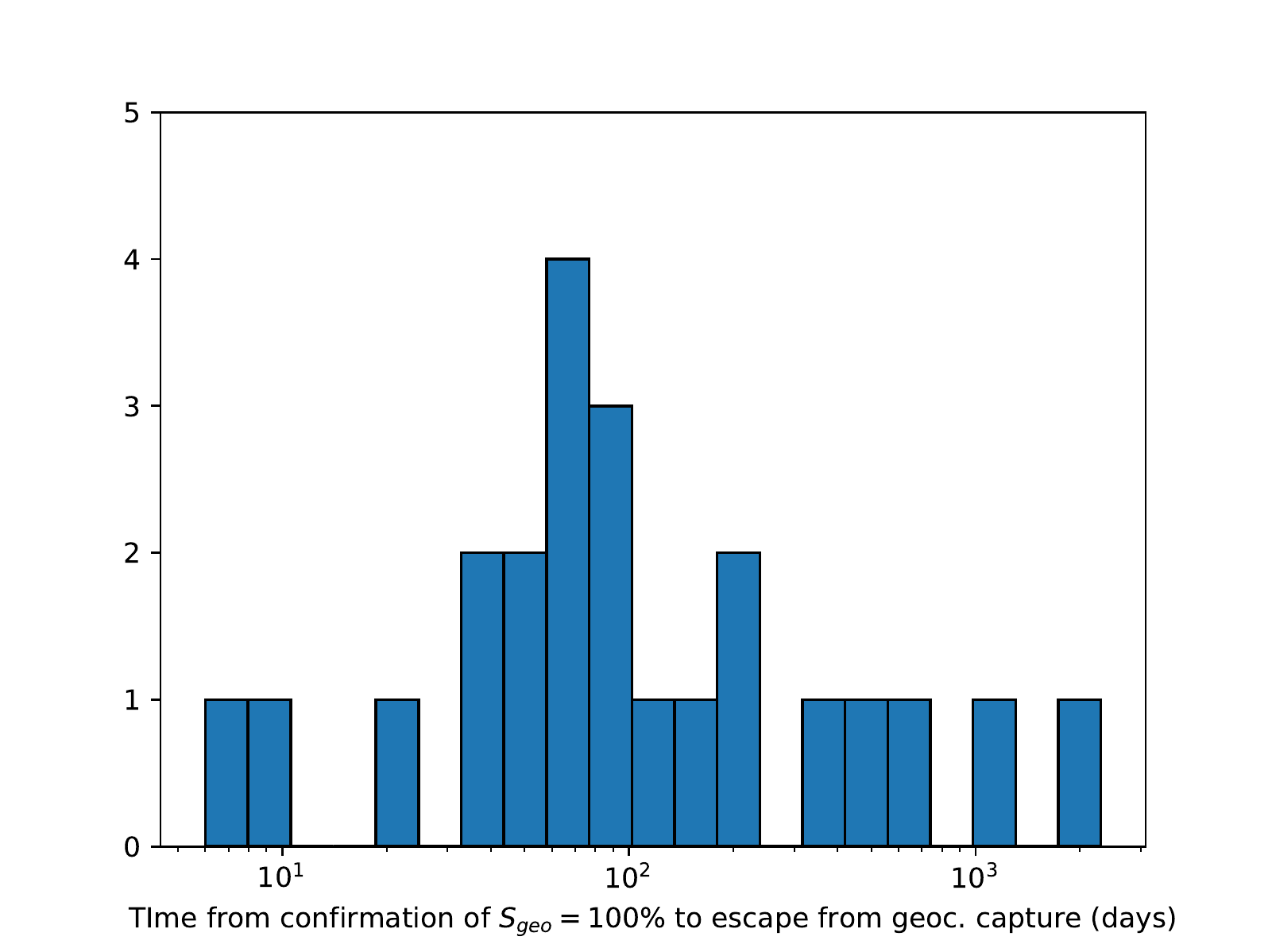}%
\caption{A histogram of the difference in time between the confirmation of the geocentric capture of an asteroid and the escape of the asteroid from the geocentric capture. The sample includes all discoverable and linkable TCOs.}%
\label{fig:r-d}%
\end{figure}

There is no specific preference for when a TCO is discovered during its capture (Fig.~\ref{fig:lineplot}). The typical duration between the confirmation of the geocentricity of the object and its release from the geocentric bind is 1--3 months, although there are occasionally exceptionally long captures present (Fig.~\ref{fig:r-d}). This constraint needs to be taken into account when planning follow-up strategies and space missions to TCOs.

\citet{bolin2014}  proposed that orbit determination would improve due to parallax and differences in velocities derived from tracklets.  Our calculations show that single-night data sets are not sufficient for the identification of geocentric objects with 100\% certainty. Assessing the impact of trailing information on the accuracy of the resulting orbits is a subject of a future study.

Aside from flyby NEAs, TCOs will be among the smallest objects detected. A characteristic feature of the TCO signal is its long trail compared to other natural objects. \rcom{Although velocities of detected objects can reach up to 1000$^{\circ}$/day, we} predict the velocity of a discoverable TCO to fall in the range from 1$^{\circ}$/day to 50$^{\circ}$/day, and to have an apparent magnitude $21 < V < 23$ (Figs.~\ref{fig:trlo} and \ref{fig:vvvel}). They are typically observable for only two days. These features are similar to those of smallest PHAs.  In their analyses, \citet{jones2018} used an upper velocity limit of 0.5$^{\circ}$/day, whereas \citet{veres2017a} used the upper limit of 2$^{\circ}$/day. As the typical velocities of TCOs range from 1$^{\circ}$/day to 100$^{\circ}$/day, the performance analysis of MOPS by these authors is not directly applicable for our purposes.

\subsection{Linking}

\rcom{Detection linking is an essential step prior to orbit computation. We assessed two linking algorithms, both of which are alternative approaches to the current linking approach in the LSST MOPS: First, linking in ephemeris space  \citep{granvik2005,granvik2008,granvik2007a}; and second, through a direct comparison of computed orbital elements \citep{granvik2007b}. With both algorithms, we linked pairs of tracklets instead of triples required by MOPS. In the following analysis, contrary to what is conventionally done in a moving object processing pipeline, the linking performance is assessed using tracklets that are a priori known to belong to the same object.}

%
%
%

For the assessment we selected the objects with $S_\mathrm{geo}=100\%$ and further required the objects to fulfill the following criteria: 

\begin{enumerate}
	\item The observations must occur during the period of geocentric capture of the object.
	\item Each object must have at least two detections on two separate nights (that is, four detections in total).
\end{enumerate}

This narrowed the selection down to 20 objects. First, the orbit solution for each tracklet was computed using two-body statistical ranging and in that process we varied the number of sample orbits (10$^4$, 10$^5$ and 10$^6$). Then, we assessed whether the ephemerides from the tracklets overlapped at three different epochs. We propagated the orbits using $n$-body integration, taking into account all planets, Pluto, and the Moon as a separate body from the Earth, as well as the relativistic correction \citep{sitarski1983}. We varied the size of the two-dimensional ($\alpha,\delta$)-cell at each selected epoch: $5^{\prime} \times 5^{\prime}$, $10^{\prime} \times 10^{\prime}$, $30^{\prime} \times 30^{\prime}$, and $60^{\prime} \times 60^{\prime}$. A linkage was deemed successful if at least one pair of ephemerides from different tracklets overlapped at three different epochs. As expected, the outcome depended on the combination of cell size and the number of sample orbits --- the ability to identify correct links increased when either increasing the number of sample orbits or the cell size. The following combinations of cell size and number of sample orbits produced the best results: $5^{\prime} \times 5^{\prime}$ cells and $10^6$ orbits, or $30^{\prime} \times 30^{\prime}$ cells and $10^5$ orbits. For completeness, we also attempted linking borderline cases ($25\%<S_\mathrm{geo}<40\%$). Also in these cases the linking was successful.

In addition to comparing address spaces at different epochs we attempted the linking of Keplerian elements. From $10^6$ orbital solutions for a tracklet we selected the orbits that would produce geocentric solutions. We then attempted to link the orbits at the mid-epoch between two tracklets. It turned out that for all geocentric objects there was a preliminary solution with a geocentric Keplerian orbit, and that it was linkable to a corresponding tracklet on the following night.

\subsection{Sources of confusion}

The typical rates of motion of TCOs imply that the majority of them will be trailed in LSST images. This is different compared to other asteroids, and even NEOs, the overwhelming majority of which will be point sources in LSST images. The maximum rate of motion for the PSF to remain pointlike is 1$^{\circ}$/day \citep{veres2017b} due to LSST's short exposure time. TCOs will thus be a rare population of objects with trailed PSFs. Other trailed objects in the LSST images that could potentially cause confusion include NEOs during close and fast flybys (such as potentially-hazardous asteroids), meteors, artificial satellites, and space debris. In what follows, we will discuss these sources of confusion in greater detail.

\subsubsection{Closely approaching asteroids}

\rcom{The combination of a rapid increase in brightness and fast velocity (cf. Fig.~\ref{fig:ratemotion}) is certainly not exclusive to TCOs, but is observed in any asteroid passing close to the Earth. Here we assess whether the closely approaching asteroids which do not become captured on a geocentric orbit may act as a source of confusion.}

\rcom{We compared TCOs to a population of closely approaching asteroids with orbital parameters in the Earth-capture zone (0.9 au $< a_{\Sun} <$ 1.1 au; $e_{\Sun} <$ 0.1; $i_{\Sun} < 2^{\circ}$) as depicted in \citet{granvik2012} and \citet{fedorets2017}. This allows us to directly compare the ratio between false positive TCOs and true TCOs.  We created synthetic astrometry for these objects for the two days around their closest Earth approach, computed orbits with statistical ranging, and estimated their $S_{geo}$. }

\rcom{In one out of 300 cases, we obtained exclusively geocentric elliptic solutions for one object, although the synthetic orbit had a minimum $e_{\Earth} = 1.02$. This implies that the object fulfilled the criteria for being, at least, a TCF. However, further assessment revealed that none of the computed orbits could be classified as a TCO. It is clear that besides computing $S_{geo}$, the orbit solution needs to be integrated back and forward in time to assess an object's true nature. Taking the above ratio of false-positive TCFs at face value and comparing it to the average capture efficiency for true TCFs of $3\times10^{-6}$, we predict that, based exclusively on LSST observations over two nights without any further follow-up observations, the false-positive-to-true ratio for TCF candidates is approximately 1000:1.
}

\subsubsection{Artificial satellites and space debris}

An object on a low-Earth orbit (LEO)  will be defocused  as has been seen in observations obtained with the Su\-ba\-ru telescope and SDSS \citep{iye2007,bektesevic2014}. Most artificial objects are in the LEO region, and will produce characteristic trails throughout the entire image \citep{schneider2012}. These characteristic observational features allow us to distinguish them from TCOs.

There are on the order of 1\,000 undisrupted, 1-10-metre sized GEO objects \citep{klinkrad2006}. Also, \citet{schneider2012} has shown that linking GEO detections is possible in theory. GEO objects will, by definition, have a distinct rate of motion of 360$^{\circ}$/day, and a geocentric eccentricity and equatorial inclination close to zero. However, TCOs do not appear on such orbits \citep{granvik2012,fedorets2017}, which allows GEOs to be filtered out based on their geocentric orbital elements. The same approach will apply to artificial satellites on graveyard GEO orbits, as their geocentric orbital elements do not differ much from those of proper GEO objects.

The highest probability of confusion is with the distant artificial satellites. These objects are usually upper stages of rocket launches away from the Earth-Moon system, towards the Moon, or towards Earth's Lagrange points $L_1$ and $L_2$. The known population of these objects are occasionally increased with the discovery of new objects and it typically takes a relatively long time until their true nature can be confirmed. For now, the best tool is calculating the area-to-mass ratio of objects from orbital deviations caused by radiation pressure \rcom{\citep[e.g.][]{micheli2013,micheli2014,jedicke2018}}. The possibility of confusion between TCOs and distant artificial satellites is therefore essentially unavoidable at discovery. For example, 2006 RH$_{120}$ was thought to be artificial before it was realised that it was indeed Earth's first discovered temporary natural satellite \citep{kwiatkowski2009}. On the contrary, 2018 RV$_{2}$ required around a month of follow-up observations before a natural origin could be ruled out for this artificial object. The number of distant artificial satellites is rather limited with 44 objects currently listed on the Distant Artificial Satellites Observations Page of the Minor Planet Center. The known distant artificial satellites can therefore be filtered out easily from the LSST data.

\subsubsection{Meteors}

There will be on average 15 meteors and fireballs in each LSST image (Peter Brown, private communication). Similarly to artificial satellites in LEO, meteor PSFs will be out of focus. Meteor PSFs will also be wider than those of artificial satellites in LEO due to the plasma cloud surrounding them during the atmospheric entry \citep{bektesevic2018}. There is thus a negligible possibility that TCOs and meteors are confused with each other.

\subsubsection{False detections}

The last source of confusion are false detections that are erroneously ingested into the MOPS pipeline. False detections include cosmic rays, stationary transient objects such as, e.g., peculiarly shaped gamma-ray bursts, other transient astronomical objects, artifacts from image subtraction caused by, e.g., seeing variations, and detector-level artifacts. As a first step, all stationary transient objects will be removed by MOPS, but some will slip through the filters.

In the worst-case scenario, each image will contain ca. 450 false detections per square degree \citep{jones2018} implying a ratio of 3:1 between false and true detections of moving objects. \citet{veres2017b} and \citet{jones2018} have shown that the successful computation of an orbit that reproduces the detections is a sufficient filter to distinguish between false and true detections.

\section{Conclusions} \label{s:c}

Based on our analyses, the baseline MOPS requirement of three observations during 15 days is too restrictive for the majority of otherwise detectable TCOs. Thus, the number of discoveries of TCOs will not provide a strong constraint on the TCO size-frequency distribution. To enable TCO detections useful for follow-up detections and prompt orbit determination, it is necessary to construct a tool complementary to MOPS to deal with  the long trails characteristic of fast flybys and TCOs.  The tool would include filtering LSST Difference Image Analysis Sources by the apparent sky velocity, PSF width and brightness; linking tracklets; producing up to $10^6$ sample initial orbits for a single tracklet; linking resulting geocentric candidates with similar candidates for the other nights; in case of a successful linkage, making a preliminary orbit determination; and, finally, in case of the success of all previous steps, making a rapid response alert, e.g. to a network of observatories. A minimum of two nights is required to make any useful predictions, even in the case of very fast moving objects. Such a system will benefit not only observations of TCOs, but also detections of metre-class PHAs. 

Current methods require at least two observations per night for initial orbit computation.  However, there are currently orbit determination methods in development for trailed astrometry \citep[Oca\~{n}a Losada \& Granvik, in preparation, P\"{o}ntinen \& Granvik, in preparation, see also ][]{virtanen2016}. These methods allow us to compute an orbit from a single streak, and to expand the treatment to single-tracklet nights, potentially doubling the number of detectable TCOs. The practical lower limit for the streak length in the streak detection method of \citet{virtanen2016} is 20 pixels, which translates to 6$^{\circ}$/day in an LSST image.

Assessing the geocentric score turned out to be a powerful tool to identify TCO candidates. \rcom{In addition, the resulting orbits of TCO candidates need to be investigated for their capture duration and number of geocentric orbits to distinguish them from TCFs. TCFs will be difficult to identify, because it is easy to confuse non-captured close-approaching objects and TCFs.}  Prompt inter-night linking and initial orbit determination allow rapid-response target-of-opportunity programmes to provide follow-up observations. The general approach for follow-up observation programmes of TCOs would be similar to alert systems common for supernovae, gamma-ray bursts, and other fields of transient astronomy. These programmes could include dedicated astrometric follow-up observations with two-metre-class telescopes, and possible physical characterisation with spectroscopy with ten-metre class telescopes. Physical characterisation is important for distinguishing the origin of TCOs from direct captures from the NEO population  on the one hand \citep{granvik2012,fedorets2017} and the lunar ejecta on the other hand \citep{gladman1995}.   This would require a number of two-metre class telescopes spread throughout the Earth for maximum fol\-low-up coverage.  

The north ecliptic spur is an essential addition to the LSST cadence for solar system objects in general. A rolling cadence would may also work well for TCOs, especially with dedicated observations in the direction of the opposition. \rcom{We estimate that the upper limit for the number of new detections would increase by 50\% compared to the basic cadence. This estimate is based on the proportion of TCOs which have high enough velocities to traverse to an adjacent field the following night.} Obtaining data separately from consecutive observations (2x15 s instead of a single 30 s integration) would reveal the direction of motion of the TCO on the sky. 

Assuming a tailored treatment of TCO data, we expect to find a TCO on average once every two or three months, resulting in the order of 75 objects  during the nominal ten-year operation period of LSST. The potential deviation from that estimate can be due to  inconsistencies in the NEO SFD in the 0.1-10 m range; uncertainty in the \citet{brown2002} slope; and smaller numbers of asteroids than predicted by population models with low $e$ and $i$ \citep{harris2015,fedorets2017}. 

Based simply on average capture duration, including TCFs might increase the detection rate by another 25\%. Another possibilities to increase TCO detections would be to relax the 100\% requirement for the geocentric score. In that case, the discovery rate of TCOs can increase by 20\%, and in some cases, even very fast single-day appearances could be within reach of follow-up observations, provided response within up to one hour is available. The combined error may shift the expected number of observations by a factor of two. The overall improvement by a tailored approach could be of one order of magnitude compared to the baseline treatment. The impressive technical capabilities of LSST allow for discovering TCOs on a regular basis.

\section*{Acknowledgements}

\rcom{We would like to thank the referees for their helpful reviews}. This work was funded by the Emil Aaltonen foundation and grants \#299543 and \#307157 from the Academy of Finland. RLJ and MJ acknowledge support from the DIRAC Institute in the Department of Astronomy at the University of Washington. The DIRAC Institute is supported through generous gifts from the Charles and Lisa Simonyi Fund for the Arts and Sciences, and the Washington Research Foundation. GF thanks Jenni Virtanen and Olli Wilkman for helpful discussions about space debris. Computational resources were provided by CSC - The Finnish Centre for Scientific Computing, Ltd.

\section*{References}

\bibliographystyle{model5-names}\biboptions{authoryear}
\bibliography{minimoons,lsst,orbit}

\end{document}